\documentclass[pre,aps,twocolumn,showpacs]{revtex4}
\usepackage{epsfig}
\usepackage{bm}
\newcommand{\B}[1]{{\bm{#1}}}
\newcommand{\C}[1]{{\mathcal{#1}}}

\newcommand{\Onecol}
{\begin{widetext}
\onecolumngrid}
\newcommand{\Twocol}
{\end{widetext} \twocolumngrid}

\newcommand{\be}{\begin{equation}}
\newcommand{\ba}{\begin{array}}
\newcommand{\bea}{\begin{eqnarray}}
\newcommand{\bfi}{\begin{figure}}
\newcommand{\ee}{\end{equation}}
\newcommand{\ea}{\end{array}}
\newcommand{\eea}{\end{eqnarray}}
\newcommand{\efi}{\end{figure}}

\newcommand{\ra}{\right\rangle}
\newcommand{\la}{\left\langle}
\begin{document}

\title{Shell Model for Drag Reduction with Polymer
Additive in Homogeneous Turbulence}
\author{Roberto Benzi$^{1}$, Elisabetta De Angelis$^{2}$,
Rama Govindarajan$^{3}$ and Itamar Procaccia$^{4,5}$}
\affiliation{$^1$ Dipartimento di Fisica and INFM, Universit\`a
``Tor Vergata",
Via della Ricerca Scientifica 1, I-00133 Roma, Italy\\
$^3$ Eng. Mech. Unit, Jawaharlal Nehru Centre for
Advanced Scientific Research, Jakkur, Bangalore 560064, India\\
$^2$ Dipartimentao di Meccanica e Aeronautica, Universit\'a di Roma La Sapienza,
via Eudossiana 18, Roma 00184, Italy\\
$^4$ Dept. of Chemical Physics, The Weizmann Institute of
Science, Rehovot, 76100 Israel\\$^5$ Dept. Of Physics, The
Chinese University of Hong Kong, Shatin, Hong Kong
.}
\begin{abstract} Recent direct numerical simulations
of the FENE-P model of non-Newtonian hydrodynamics revealed that
the phenomenon of drag reduction by polymer additives exists
(albeit in reduced form) also in homogeneous turbulence. We
introduce here a simple shell model for homogeneous viscoelastic
flows that recaptures the essential observations of the full
simulations. The simplicity of the shell model allows us to offer
a transparent explanation of the main observations. It is shown
that the mechanism for drag reduction operates mainly on the
large scales. Understanding the mechanism  allows us to predict
how the amount of drag reduction depends of the various
parameters in the model. The main conclusion is that drag
reduction is not a universal phenomenon, it peaks in a window of
parameters like Reynolds number and the relaxation rate of the
polymer. \vskip 0.2cm
\end{abstract}
\maketitle

\section{Introduction}
The phenomenon of drag reduction by polymer additives is usually
studied in channels or pipes, where the boundary conditions and
the effects of the walls are very important
\cite{69Lum,75Vir,87LT,97THKM}. Until recently it was not known
whether drag reduction can be achieved also in homogeneous flows;
this question has been answered recently in the affirmative, via
Direct Numerical Simulations (DNS) of the FENE-P model equations
\cite{87BCAH,94BE} in homogeneous conditions (i.e. in a box with
periodic boundary conditions) \cite{03ACBP}. The FENE-P model
takes the effect of the polymers on the Newtonian fluid into
account by introducing the conformation tensor $\B R$ of the
polymers into the fluid stress tensor. The FENE-P equations are
known to model well the effects of polymers on the hydrodynamic
flows, and DNS of these equations in channel geometry recaptured
very well the characteristics of drag reduction in experimental
channel turbulence \cite{97THKM,02ACP}. The observation of drag
reduction in homogeneous conditions offers an opportunity to
investigate the phenomenon independently of boundary layers and
wall effects. Nevertheless the FENE-P equations are relatively
cumbersome to analyze without the help of DNS. The aim of this
paper is to introduce a shell model of the homogeneous FENE-P
equations. We will demonstrate that the shell model recaptures
the main findings of the homogeneous DNS, and that these findings
are understandable analytically, taking advantage of the relative
simplicity of the shell model. To derive the shell model for drag
reduction we make use of a formal analogy between the FENE-P
equations for viscoelastic flows and magneto-hydrodynamics (MHD)
\cite{01BFL}. It had been pointed out that if we form a tensor
$R_{i,j}$ from the direct product of the magnetic field $B_i$,
i.e. $R_{i,j}\equiv B_i B_j$, then the nonlinear couplings of MHD
lead to equations for the tensor $\B R$ whose nonlinear terms are
equivalent to those of FENE-P, up to terms that remove the dynamo
effect. This analogy is revisited and exploited in Sect.
\ref{analogy}. The shell model for viscoelastic flow is
introduced and discussed in Sect. {\ref{shell}}. In Sect.
\ref{simulations} we present numerical simulations of the shell
model and demonstrate the existence of drag reduction. In Sect.
\ref{mechanism} we present the mechanism of drag reduction. This
is the central section of this paper. We show that drag reduction
is not a universal phenomenon. Rather, it depends on the
parameters, like the Reynolds number and the relaxation time of
the polymer. The amount of drag reduction peaks in a window of
these parameters. In Sect. VI we demonstrate that understanding
the mechanism provides us with predictive power that we can test
against numerical simulations. We conclude in Sect.
\ref{conclusions} by observing that precisely because drag
reduction is not a universal phenomenon it can be manipulated by
optimizing parameters.

\section{The FENE-P equations and their relation to MHD}
\label{analogy} The addition of a dilute polymer to a Newtonian
fluid gives rise to an extra stress tensor $\B {\C T}(\B r,t)$
which affects the Navier-Stokes equations \cite{87BCAH,94BE}:
\begin{eqnarray}
\frac{\partial \B u}{\partial t}+(\B u\cdot \B \nabla) \B u&=&-\B \nabla p
+\nu_{\rm s} \nabla^2 \B u +\B \nabla \cdot \B {\C T}\ , \nonumber\\
\B \nabla \cdot \B u &=& 0 \ . \label{Equ}
\end{eqnarray}
Here $\B u(\B r,t)$ is the solenoidal velocity field,  $p(\B r,t)$ is the pressure
and $\nu_s$ is the viscosity of the neat fluid.
In the FENE-P model the additional stress tensor $\B {\C T}$ is
determined by the ``polymer conformation tensor" $\B R$ according to
\begin{equation}
\B {\C T}(\B r,t) = \frac{\nu_p}{\tau}
\left[\frac{f(\B r,t)}{\rho_0^2} \B R(\B r,t) -\B 1 \right] \ .
\end{equation}
Here $\B 1$ is the unit tensor, $\nu_p$ is a viscosity parameter,
$\tau$ is a relaxation time for the
polymer conformation tensor and $\rho_0$ is a parameter which in the
derivation of the model stands for the rms extension of the
polymers in equilibrium. The function $f(\B r,t)$ limits the growth of the trace of
$\B R$ to a maximum value $\rho_{\rm m}$:
\begin{equation}
f(\B r,t) \equiv \frac{\rho_{\rm m}^2-\rho_0^2}
{\rho_{\rm m}^2 -R_{\gamma\gamma}(\B r ,t)} \ .
\end{equation}
The model is closed by the equation of motion for
the conformation tensor which reads
\begin{eqnarray}
\frac{\partial  R_{\alpha\beta}}{\partial t}&+&(\B u\cdot \B \nabla) R_{\alpha\beta}
=\frac{\partial u_\alpha}{\partial r_\gamma}R_{\gamma\beta}
+R_{\alpha\gamma}\frac{\partial u_\gamma}{\partial r_\beta}\nonumber\\
&-&\frac{1}{\tau}\left[ f(\B r,t)  R_{\alpha\beta} -\rho_0^2 \delta_{\alpha\beta} \right] \ .
\label{EqR}
\end{eqnarray}
This model was simulated by DNS in channel flow turbulence,
showing qualitative and quantitative agreement with laboratory
experiments on drag reduction. Recently the same model has been
used to understand whether or not drag reduction is observed in
homogeneous and isotropic conditions \cite{03ACBP}. In homogeneous
and isotropic turbulence, drag reduction can be determined by
computing the ratio
\begin{equation}
D = \frac{\epsilon L}{E^{3/2}} \ ,
\end{equation}
where $E$ is the kinetic energy,$\epsilon$ is the total rate of
energy dissipation and $L$ is the scale of the external forcing.
The above expression of drag reduction can be easily reduced to
the so called skin friction factor for turbulent channel flows.
The numerical simulation of homogeneous and isotropic turbulence
were performed in a cube with periodic boundary conditions. The
external forcing was applied with random phase in order to ensure
isotropy and homogeneity. The numerical simulations were
performed for the Navier-Stokes equations and the FENE-P model
for the {\em same} external forcing. Both the total energy
dissipation and the kinetic energy increased for the FENE-P as
compared to the Newtonian case. A direct computation of $D$ shows
that there is a drag reduction of about $20\%$, i.e. roughly of
the same order as what had been observed in turbulent channel
flow. Also, in homogeneous and isotropic turbulence, the Taylor
microscale appeared to increase, apparently precisely as much as
the buffer layers increases in channel flows \cite{69Lum}. This is
an interesting result because it tells us that the effect of
boundary conditions is not crucial for drag reduction, at least
from a physical point of view. Nevertheless, it is still
difficult to understand from numerical simulations, even in the
homogeneous and isotropic case, what is the physical mechanism
that is responsible for drag reduction. The increase of the
Taylor microscale is certainly not enough to explain
quantitatively the increase of the kinetic energy, as somehow
previously suggested in the literature \cite{69Lum,90Gen}.

Having understood that the homogeneous simulations exhibit drag
reduction, we would like to propose a mechanism for it. Rather
than doing it directly with the FENE-P model, we would present
first a simplified model. We have already shown before that drag
reduction appears in simplified models like the Burgers equation
\cite{03BP}. Here we derive a shell model for the FENE-P
equations. The advantage of the shell model is that it is much
more tractable analytically than the full FENE-P equations. We
will present the model, demonstrate explicitly that it exhibits
drag reduction in much the same way as the FENE-P equations, and
finally offer a new mechanism to understand the phenomenon.

\section{The shell model}
\label{shell} To derive a shell model of the homogeneous FENE-P
equations (without boundaries) we proceed in two steps. First we
recall a recent remark \cite{01BFL} that the FENE-P equations can
be recaptured almost entirely by taking the conformation tensor to
be a diadic direct product of of a vector $\B B$, i.e
$R_{ij}\equiv B_i B_j$. In terms of this vector the equations read
\begin{eqnarray}
\frac{\partial \B u}{\partial t}+(\B u\cdot \B \nabla) \B u&=&-\B \nabla p
 +\B B \cdot (\nabla \B B)
+\nu_{\rm s} \nabla^2 \B u\ , \nonumber\\
\B \nabla \cdot \B u &=& 0 \ , \nonumber\\
\frac{\partial \B B}{\partial t}+(\B u\cdot \B \nabla) \B B&=&
-\frac{\B  B}{\tau} +\B B \cdot (\nabla \B u) \ , \nonumber\\
\B \nabla \cdot \B B&=& 0  \ . \label{Bmodel}
\end{eqnarray}
These equations are identical to the FENE-P model up to the
explicit appearance of the function $f(\B r,t)$. The learned
reader of course recognizes that for $\tau\to \infty$ these
equations are isomorphous to magneto-hydrodynamics (MHD). We can
therefore write immediately, by inspection, a shell model for
FENE-P by using the well studied shell model for MHD
\cite{98GC,03CCGP}, including the relaxation term for finite
$\tau$. We denote the velocity field by $u$, and the ``polymer "
field by $B$. The dynamical variable of the shell model are the
field at wave vector $k_n$, denoted respectively as $u_n\equiv
u(k_n)$ and $B_n\equiv B(k_n)$. The shell model restricts
attention to wavevectors $k_n=k_0\lambda^n$, where typically in
numerical simulations $\lambda=2$.

In order to derive the shell model equation,
we consider the following non linear operator:
\bea
\Phi_n(u,B) = b_1k_nu_{n+2}B^*_{n+1}-b_1k_{n-1}u_{n+1}B^*_{n-1} + \nonumber\\
c_1k_nu^*_{n+1}B_{n+2}+c_1k_{n-2}u_{n-1}B_{n-2} + \nonumber\\
b_ck_{n-1}B_{n+1}u^*_{n-1}+b_ck_{n-2}u_{n-2}B_{n-1} \eea where
$b_1 = 1-b;$ $ c_1 = 1+c;$ $ b_c = b+c$, and $-1\le b\le 0$ and
$c= 1+b$ are the usual parameters defined in the Sabra model.

In terms of the non linear operator $\Phi$, the Sabra shell model
of turbulence \cite{98LPPPV} can be written as \be \frac{du_n}{dt}
= \frac{i}{3} \Phi_n(u,u) - \nu k_n^2 u_n + f_n \label{sabra} \ee
where $f_n$ is an external forcing and $\nu$ the kinematic
viscosity of the model. Let us remark that the following relation
can be proved: \be i\Sigma_n \Phi_n(u,B) B^*_n - i\Sigma_n
\Phi_n^*(u,B) B_n = 0. \label{sigma} \ee

Using the non linear operator $\Phi$ it is possible to model
equations (\ref{Bmodel}) in the framework of shell models, namely:

\begin{eqnarray}
\frac{du_n}{dt} = \frac{i}{3} \Phi_n(u,u) -\frac{i}{3}\Phi_n(B,B)
-  \nu k_n^2 u + f_n \nonumber \\
\frac{dB_n}{dt} = \frac{i}{3} \Phi_n(u,B) -
\frac{i}{3}\Phi_n(B,u) - \frac{1}{\tau} B_n \ . \label{sabrab}
\end{eqnarray}
Equation (\ref{sigma})  tells us that the generalized energy $E$,
\begin{eqnarray}
E &=& E_u +E_B \nonumber\\
E_u&\equiv&\Sigma_n u_nu^*_n \ , \quad E_B\equiv \Sigma_n
B_nB^*_n \ . \label{defE}
\end{eqnarray}
is conserved in the inviscid limit, i.e. for $\tau \rightarrow
\infty$ and $\nu \rightarrow 0$.

We will refer to this model as the SabraP model. Beside the
generalized energy $E$, the model conserves the ``cross helicity''
in the inviscid limit \be K = \sum_n \Re(u_n^* B_n).
\label{crosshe} \ee

In MHD one needs to worry about the existence of a dynamo effect,
i.e. an unbounded increase in the magnetic field. In our case the
term that models the polymer relaxation time $-B_n/\tau$ will be
responsible for guaranteeing stationary statistics without dynamo.
In addition to the conservation laws the equations of motion
remain invariant to the phase transformations $u_n\to
u_n\exp(i\phi_n)$ and $B_n\to B_n\exp(i\psi_n)$. The conditions
are
\begin{eqnarray}
\phi_n + \phi_{n+1} - \phi_{n+2} &=& 0\ ,
\label{phases1}\\
\phi_{n} + \psi_{n+1} - \psi_{n+2} &=& 0\ ,
\label{phases2}\\
\psi_n + \phi_{n+1}-\psi_{n+2} &=&0\ ,
\label{phases3}\\
\psi_{n} + \psi_{n+1} - \phi_{n+2} &=& 0\ .
\label{phases4}
\end{eqnarray}
This implies $\psi_n = \phi_n$ $\forall n$.
As a result of the phase constraints there exist in this model
only few non-zero correlation functions. The only second order
quantities are $\langle |u_n|^2\rangle$ and $\langle |B_n|^2\rangle$.
The only third order quanitites are of the form
$\Im \langle \beta_{n-1}\beta_n \beta^*_{n+1}\rangle$ where $\beta$ can be $u$ or $B$.

\section{Numerical investigation of the shell model: drag reduction}
\label{simulations}

In this section we compare the solutions of the shell model (\ref{sabrab}) to the
usual Sabra shell model for the corresponding Newtonian flow.
The Sabra model (\ref{sabra}) is obtained from (\ref{sabrab}) in the limit $\tau \to 0$.
Alternatively, we can get the Sabra dynamics by simply taking as
initial conditions $B_n=0$.

To have a meaningful comparison we always drive the two models
with a constant power input. In other words, we choose \be f_1 =
\frac{F_a}{u^*_1} \;\;\; f_2 = \frac{F_b}{u^*_2} \;\;\; f_n = 0
\;\; for \;\; n\ge2 \ , \ee with $F_a = F_b = 10^{-3} (2+2i)$.
Since the power input is the same, drag reduction is exhibited in
(\ref{sabrab}) if the kinetic energy of the flow increases. The
latter is simply  $ \langle E_u\rangle $. We will investigate the
existence of drag reduction, its dependence on parameters, the
question of the dissipative scale, and the dynamical signatures of
drag reduction.
\subsection{Drag reduction and its dependence on parameters}
\begin{figure}
\centering
\includegraphics[width=.5\textwidth]{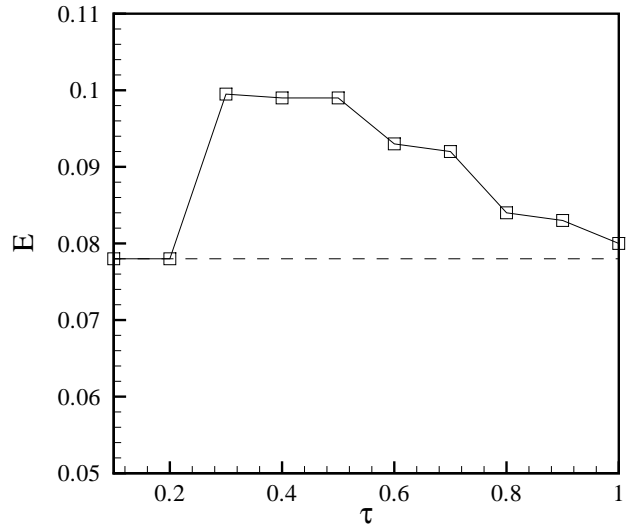}
\caption{Kinetic energy of the SabraP model for $\nu= 10^{-5}$ as
a function of $\tau$. The constant reference line corresponds to
the kinetic energy computed for the Sabra model without polymer.}
\label{fig1}
\end{figure}

\begin{figure}
\centering
\includegraphics[width=.5\textwidth]{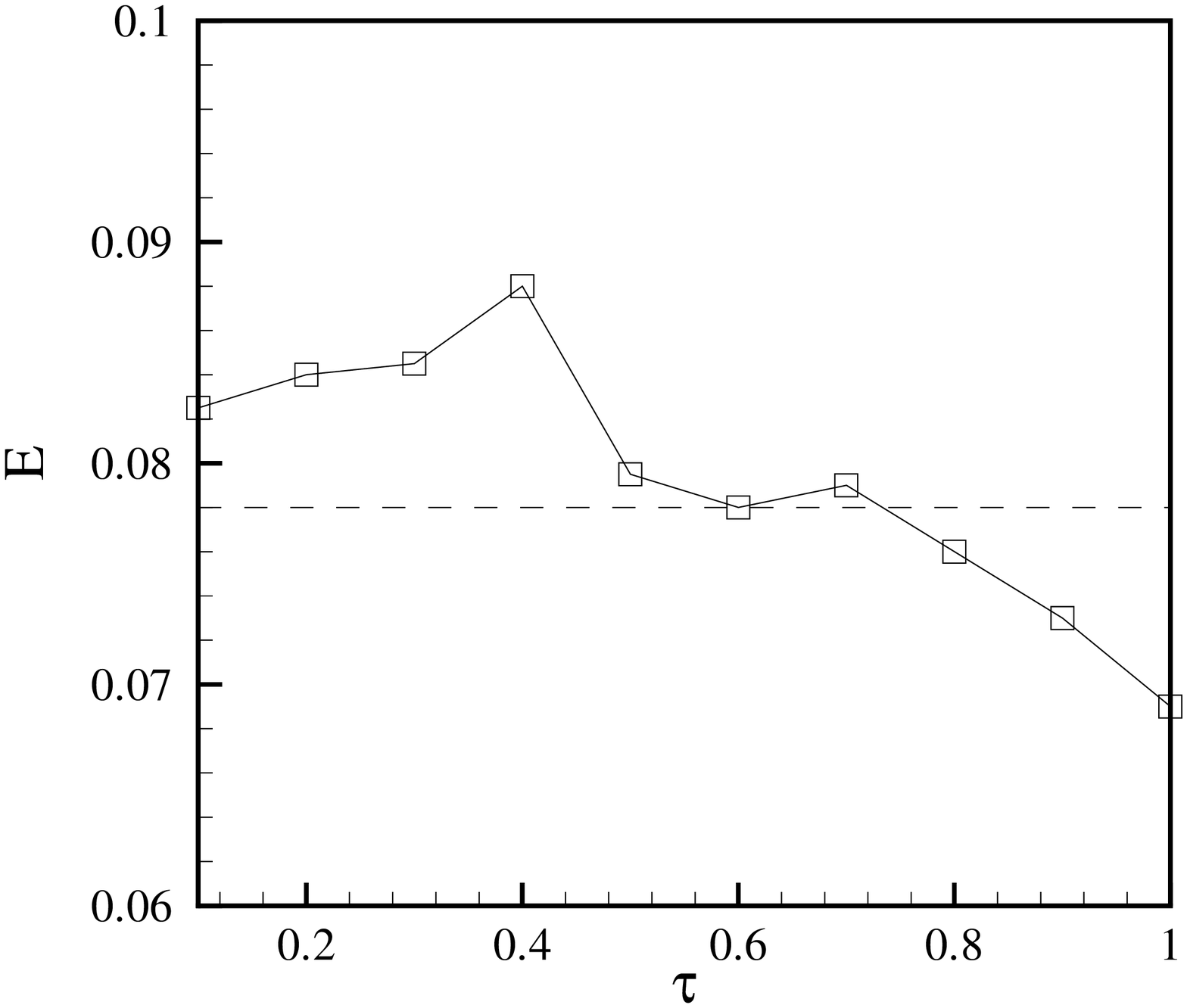}
\caption{Kinetic energy of the SabraP model for $\nu= 10^{-6}$ as
a function of $\tau$. The constant reference line corresponds to
the kinetic energy computed for the Sabra model without polymer.}
\label{fig2}
\end{figure}

\begin{figure}
\centering
\includegraphics[width=.5\textwidth]{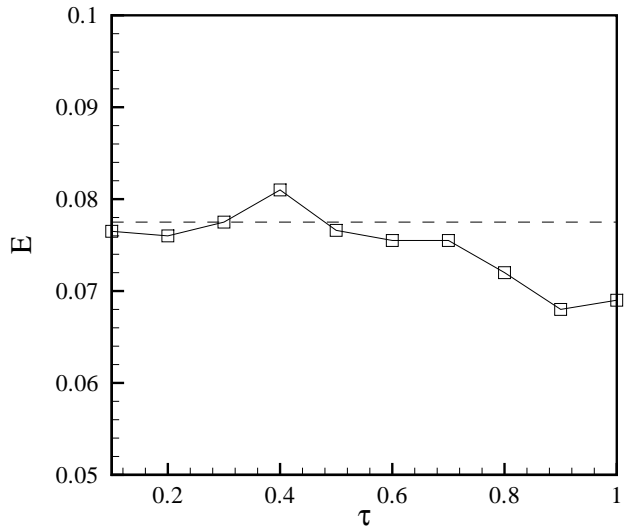}
\caption{Kinetic energy of the SabraP model with polymer for
$\nu= 10^{-8}$ as a function of $\tau$. The constant reference
line corresponds to the kinetic energy computed for the Sabra
model without polymer.} \label{fig3}
\end{figure}

We have numerically investigated the behaviour of the SabraP model
for different values of $\tau$ and $\nu$. In figures
(\ref{fig1})-(\ref{fig3}) we show $\langle E_u \rangle$ for three
values of the viscosity and for different values of $\tau$. For
concreteness we have fixed the model parameter $b$ to be $-0.4$
in all the simulations. In all the figures the constant line
corresponds to the value of the kinetic energy computed for the
Sabra model without coupling to $B_n$.
By inspecting the three figures, one can safely state that the
SabraP model shows drag reduction. In particular, for all cases,
there is an optimal choice of $\tau$ for which the effect of drag
reduction is maximal. For $\tau \to 0$ and $\tau \to \infty$ drag
reduction decreases and eventually we enter a region of parameters
where we observe drag enhancement. Moreover, for fixed value of
$\tau$ and decreasing  values of $\nu$, drag reduction decreases,
reaching a mere few per cents for $\nu=10^{-8}$.
\subsection{Which scales are responsible for drag reduction?}

\begin{figure}
\centering
\includegraphics[width=.35\textwidth]{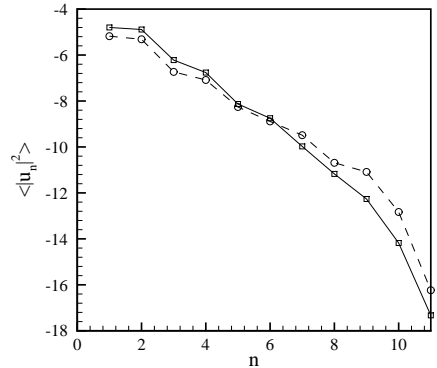}
\caption{A comparison of the average energy shell-by-shell for
the Sabra (dotted line with circles) and the SabraP (continuous
line with squares) models. Drag reduction is seen is the relative
increase in energy for small values of $n$ on the expense of
large values of $n$.} \label{spectrum}
\end{figure}

To understand which scales are responsible for the
drag reduction, we compare $\langle |u_n|^2 \rangle$
for both models with $\nu=10^{-5}$,
again at the same power input. This is shown in Fig. \ref{spectrum}.
This figure teaches us an interesting and important lesson. It is
clear that the drag reduction is due to the relative increase in
$\langle |u_n|^2 \rangle$ {\em for small values of $n$}, and that
this occurs on the expense of a relative decrease in $\langle
|u_n|^2 \rangle$ for high values of $n$. This finding is in close
correspondence with similar conclusions obtained for the FENE-P
model, both in homogeneous and channel flows \cite{03ACLPP}.
\subsection{The dissipative scale}

\begin{figure}
\centering
\includegraphics[width=.35\textwidth]{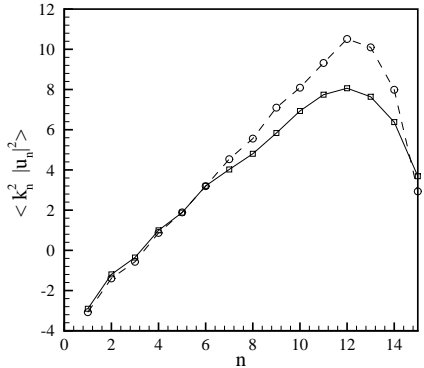}
\caption{Energy dissipation computed for each shell for the Sabra
model (dotted line with circles) and for the SabraP model
(continuos line with squares)} \label{diss}
\end{figure}

In some theories of drag reduction it was proposed that the
dissipative scale is increased in the viscoelastic flow, and that
somehow this is responsible for the phenomenon
\cite{69Lum,90Gen,00SW}. To test this possibility we plot in Fig.
\ref{diss} the quantity $\langle k^2_n |u_n|^2\rangle$ as
function of $n$.
This quantity peaks at the dissipative scale, i.e. the Kolmogorov
scale. Inspecting figure (\ref{diss}) teaches us that the
dissipative scale has not changed at all between the Sabra and
the SabraP models, even though the latter certainly exhibits drag
reduction. Thus, as indicated before, drag reduction should be
understood as a phenomenon of the energy containing scales rather
than the dissipative scales.
\subsection{Dynamical signature of drag reduction}

The similarity between the FENE-P and its shell analog transcends
statistical quantities. To observe the close dynamical similarity
it is instructive to consider the quantity
\begin{equation}
\Pi = i\Sigma_n u^*_n \Phi_n(B,B)-i\Sigma_n u_n \Phi^*_n(B,B) \ ,
\end{equation}
which describes the exchange between the kinetic energy $E_u$ and
the ``polymer'' or ``elastic'' energy $E_B$.
\begin{figure}
\centering
\includegraphics[width=.35\textwidth]{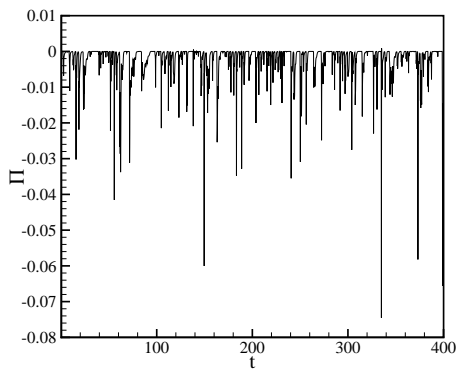}
\caption{Time behavior of the quantity $\Pi$ as defined in the
text} \label{sp}
\end{figure}
In figure (\ref{sp}) we show a time series of $\Pi$ for $\nu =
10^{-5}$. $\Pi$ is always negative; the effect of the ``polymers"
is to drain energy form the kinetic energy. Moreover the dynamics
of $\Pi$ is strongly intermittent which is a feature already
observed in the DNS of the FENE-P model. The numerical
simulations indicate the conclusion that the model introduced in
this paper shows drag reduction in a way qualitatively close to
the observed behaviour of the FENE-P model \cite{02ACP}.

We note in passing that it is not the first time that shell models
seem to reproduce many of the features of turbulent flows; it is
gratifying however that we can present a similar success even when
we include relatively non trivial effects induced by polymer
dynamics.

\section{Mechanism for drag reduction}
\label{mechanism}

This is the central section of this paper, in which we propose a
detailed mechanism for drag reduction in the present model. We
begin by analyzing the necessary conditions for drag reduction.

\subsection{Necessary condition for drag reduction}

\begin{figure}
\centering
\includegraphics[width=.5\textwidth]{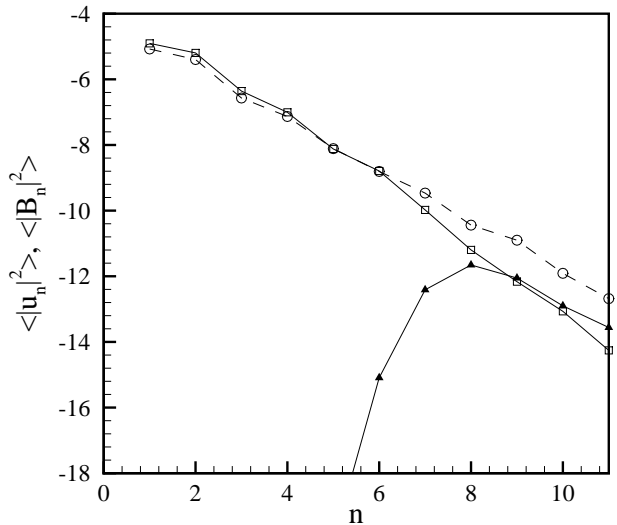}
\caption{Energy spectrum of the SabraP model (continuos line with
squares) and the Sabra model (continuos line with circles) for
$\nu=10^{-6}$. The continuos line with black triangles represents
the energy spectrum of the $B$ field.} \label{logspectrum}
\end{figure}

To derive a necessary condition for drag reduction, let us
consider the equation for the total energy, which reads:

\begin{eqnarray}
&&{\rm d} E /{\rm d} t = \Sigma_n \Big[\frac{1}{2}\left(f_nu_n^* + f_n^*u_n\right) -
\nu k_n^2
u_nu_n^* \nonumber\\&&- {1 \over \tau}B_nB_n^* \Big] \ .\label{tot-en}
\end{eqnarray}
At steady state, with power input maintained constant at $P$, we
have \be P =  \Sigma_n \left[\nu k_n^2 u_nu_n^* + {1 \over
\tau}B_nB_n^* \right] \label{tot-dissip} \ee All the terms on the
RHS are strictly positive. Since the energy input $P$ is constant
for the Sabra and the SP models, we get \be \Sigma_n
k_n^2\left[(u_nu_n^*)_S - (u_nu_n^*)_{SP}\right] > 0.
\label{diss-ineq} \ee On the other hand if the SP model is to be
drag reducing, we must have \be \Sigma_n \left[(u_nu_n^*)_S -
(u_nu_n^*)_{SP}\right] < 0. \label{drag-red} \ee The only way
(\ref{diss-ineq}) and (\ref{drag-red}) can hold simultaneously is
if for small $k_n$, $|u_n|_{SP}>|u_n|_S$, and sufficiently larger
to compensate for the fact that at large $k_n$,
$|u_n|_{SP}<|u_n|_S$. This means that the kinetic energy plotted
versus $k$ has to display an increased slope {\em at least
somewhere} for drag reduction to take place. We have seen this
already in Fig. \ref{spectrum}. We show this important phenomenon
once more in a log-log plot in Fig. \ref{logspectrum}, in which
also the $B_n$-spectrum is shown for future reference.We see very
clearly the crossing that occurs between the $u_n$ spectrum of the
SabraP model and the Sabra counterpart, which is the necessary
condition for drag reduction.  Note that the increase in slope is
a necessary but not a sufficient condition for drag reduction. We
may increase the slope but not enough to cross the Sabra
spectrum, or cross but not far enough to compensate for the
reduced kinetic energy at large $k$.
\subsection{Typical scales related to the polymer}
\label{typical}
 A discussion of the mechanism of drag reduction
calls for pointing out the existence of two typical scales that
were already introduced in the past in the literature on drag
reduction. The first is the Lumley scale, $k_c$, which is defined
by the relaxation time of the polymer being of the same order as
the eddy turn over time. For our model this scale satisfies
\begin{equation}
u(k_c) k_c \sim \tau^{-1} \ . \label{kc}
\end{equation}
Note that by definition this scale is Reynolds number independent.

\begin{figure}
\centering
\includegraphics[width=.5\textwidth]{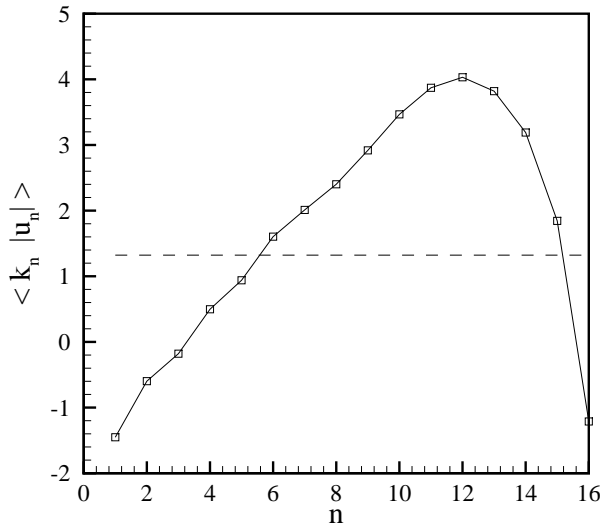}
\caption{The inverse of the ``eddy turn over time" $\sqrt{\langle
E_n\rangle }k_n$ as a function of $k_n$ (continuous line with
squares). The constant reference dotted line is $\tau^{-1}$. The
crossing in inertial range identified $k_c$. In this figure
$\nu=10^{-5}$, $\tau=0.4$.} \label{kappac}
\end{figure}

The other scale, that we refer to as the de Gennes scale $k_g$,
is where the kinetic energy on the scale $k_g$ is of the same
order as the elastic energy:
\begin{equation}
u^2(k_g) \sim B^2(k_g) \ . \label{defkg}
\end{equation}
In fact, in the SabraP model the scales so defined appear to be
very close, if not identical to each other. In particular, we will
show presently that also $k_g$ is Reynolds number independent. To
demonstrate the equivalence of the two scales we first exhibit in
Fig. \ref{kappac} the numerical estimate of $k_c$.
The physical significance of $k_c$ is not in the accidental
identity of two time scales, but rather that for $k$-vectors
smaller than $k_c$ the effect of the $B_n$ field on the energy
flux is negligible, but not so for $k$-vectors larger than $k_c$.
To see this introduce two quantities related to the energy flux
in the SabraP model, namely: \bea \label{sn}
S_n = \la Im( u_{n-1}^*u_n^*u_{n+1})\ra \\
\label{tn} T_n = \la Im( B_{n-1}^*u_n^*B_{n+1})\ra \eea The
physical meaning of the two quantities is rather clear: $S_n$
describes the flux of kinetic energy from large scale to small
scales due to non linear terms, while $T_n$ describes the flux of
kinetic energy to the polymer field. We expect that for $k_n$
near $k_c$, the effect of $T_n$ cannot be neglected in the
dynamics, i.e. the average energy flux for the velocity field
$G_n = S_n-T_n$ begins to change with respect to what it is
observed in the Sabra model.

\begin{figure}
\centering
\includegraphics[width=.5\textwidth]{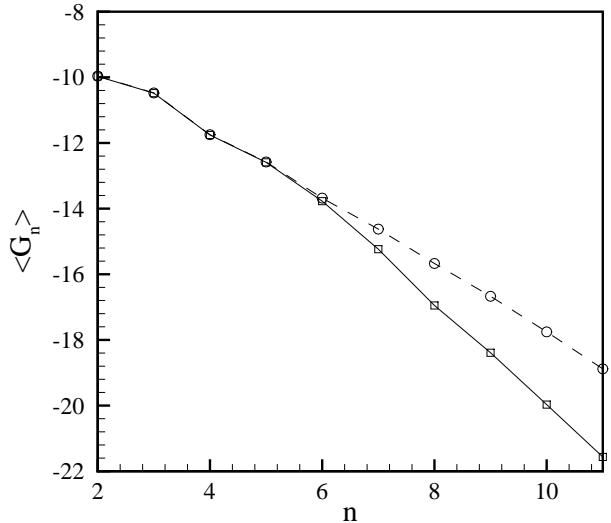}
\caption{The average energy flux $G_n$ computed for the SabraP
model (continuous line with squares) and the Sabra model (dotted
lines with circles).} \label{fluxt}
\end{figure}

In figure (\ref{fluxt}) we show the quantity $G_n$ computed for
the same model parameters of figure (\ref{kappac}). The symbols
refer to the Sabra model while the continuous line corresponds to
the SabraP model. In the vicinity of $n_c \sim 5.5$ the two models
show a different behaviour and in particular the SabraP model
shows a decrease of the total energy flux $G_n$ as previously
claimed.

\begin{figure}
\centering
\includegraphics[width=.5\textwidth]{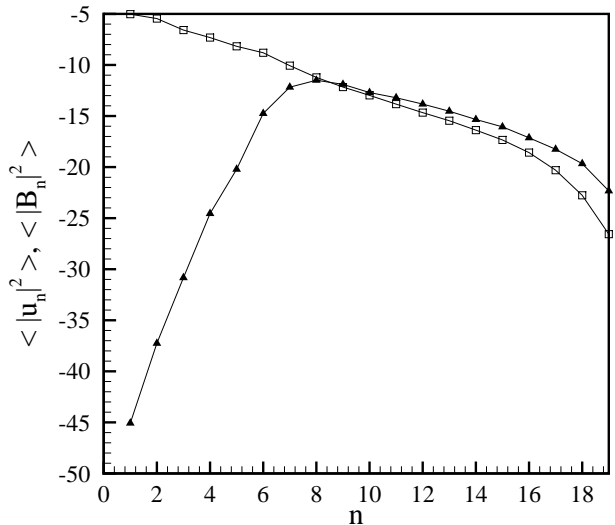}
\caption{Energy spectrum of the SabraP model (continuous line
with squares). The line with black triangles represents the
energy spectrum of the $B$ field.} \label{spre8}
\end{figure}

Regarding the scale $k_g$, it can be read from the spectrum shown
in Fig. \ref{logspectrum}, in which $\nu=10^{-6}$. In Fig.
\ref{spre8} we show the analogous spectra for $\nu=10^{-8}$.
Clearly $k_g$ did not change at all, in agreement with our
assertion that it is Reynolds independent. Finally, we note that
in all the figures shown $k_c$ and $k_g$ are of the same order of
magnitude, and in the sequel we do not distinguish between the
two.
\subsection{The effect of the polymer at large $k$-vectors}

In this subsection and the next we discuss the effect of the $B_n$
field on the $u_n$ field for $k$-vectors much larger and much
smaller than $k_c$. We will show that the spectrum $\langle
|u_n|^2 \rangle$ exhibits essentially the same scaling exponent
as the Sabra model, but the {\em amplitude} is affected by the
presence of the $B_n$ field. This will be an important ingredient
in the mechanism of drag reduction.

Begin with $k_n$ large, $k_n\gg k_c$. In this regime the effect
of the relaxation time $\tau$ on the dynamics of the $B_n$ field
is completely negligible. The dynamics of $B_n$ is dominated by
its coupling to $u_n$, simply because $u_n k_n \gg \tau^{-1}$. But
then in this regime the dynamics is like the one of MHD which had
been analyzed in detail in \cite{03CCGP}. The central conclusion
of that analysis is that up to intermittency corrections, the
spectra of both the $B_n$ and the $u_n$ fields exhibit a scaling
exponent $\zeta_2=2/3$. Indeed, inspecting Fig. \ref{spre8}, we
see that for large $k_n$ the two spectra have similar slopes,
although intermittency affects the two spectra in a different way.

On the other hand, the amplitudes of the two spectra need not be
the same.  The relative displacement of the two  power laws is
determined by numerical details in the model. To estimate this
displacement we will estimate the amplitudes of the two spectra
at the dissipative scale. The contribution to the dissipation of
$u$ is mainly from the small scales, i.e. very large values of
$k_n$. We can define an effective scale $k_d$ the scale at which
energy dissipation peaks:
\begin{equation}
\Sigma k_n^2 \langle |u_n|^2 \rangle  \sim k_d^2\langle |u_d|^2
\rangle \label{diss2}
\end{equation}
where $u_d = u(k_d)$, and $k_d$ is of the order of the Kolmogorov
scale. Since we found that the dissipative scale hardly changes
when we add the coupling to the $B_n$ field, we can deduce from
(\ref{tot-dissip}) that
\begin{equation}
\langle |u_d|^2 \rangle_{SP} - \langle |u_d|^2 \rangle_S \approx
\sum_n \frac{\langle |B_n|^2\rangle}{\nu\tau} \ .
\label{disp-diff}
\end{equation}
On the other hand, the sum on the RHS of Eq. (\ref{disp-diff}) is
a geometric sum dominated by the contribution of $B(k_c)$ where
$B_n$ is maximal. We thus estimate the relative displacement of
the two spectra at high values of $k_n$ by
\begin{equation}
\langle |u_d|^2 \rangle_{SP} - \langle |u_d|^2 \rangle_S \approx
\frac{\langle |B(k_c)|^2\rangle }{\nu\tau} \ . \label{dispdiff}
\end{equation}
Thus to first approximation we expect the slopes of the two
spectra to remain unchanged, maintained at a constant difference
from each other as given by (\ref{dispdiff}), until $k_n$
approaches $k_c$ from above, where the effect of the relaxation
time $\tau$ on the dynamics of $B_n$ cannot be neglected.

\subsection{The effect of the polymer at small $k$-vectors}

Next we discuss the slope of the $u_n$ spectrum for $k_n\ll k_c$.
This is very easy, since the amplitude of $B_n$ is very small due
to the very efficient exponential damping by $\tau$. Thus the
$u_n$ field hardly feels the coupling to $B_n$, and its slope, up
to intermittency corrections, is again of the order of
$\zeta_2=2/3$. Again, the amplitude {\bf is} changed compared to
the pure Sabra case, and this is the most important feature that
is discussed next.
\subsection{The tilt in the spectrum at $k_n\approx k_c$}

Considering the spectrum in Fig. \ref{spre8} we note that the
$B_n$ spectrum increases rapidly when $k_n\to k_c$ from the left.
To understand this phenomenon consider the equation of motion for
$B_n$ at steady state. To leading order \be 0 = \langle |{{\rm d}
B_n \over {\rm d} t}|\rangle \sim \langle|c_1 k_n u^*_{n+1}B_{n+2}
|\rangle - {\langle |B_n| \rangle\over \tau} \ , \label{simp-b}
\ee where we have neglected terms of the order of $B_{n+1}$, but
including them will lead to similar conclusions. Using the fact
that $|k_{n+1} u^*_{n+1}\tau|\ll 1$, and since $
|k_{n+1}u^*_{n+1}B_{n+2}| \le |k_{n+1}u^*_{n+1}||B_{n+2}|$ we
immediately conclude that
\begin{equation}
\langle |B_n|\rangle \ll \langle | B_{n+2}|\rangle \ .
\end{equation}
We continue this argument recursively to estimate the largest
polymer contribution $B(k_c)$ as \be \langle|B(k_c)|\rangle \sim
{\langle|B_0|\rangle \over \langle|k_1u_1|
\rangle\langle|k_3u_3|\rangle...\langle|k_{c-1}u_{c-1}|\rangle
\tau^{n_c/2}} \label{bj} \ee where $\lambda^{n_c} = k_c$.

In the vicinity of the scale $k_c$, we have to leading order
in $B$ in the kinetic energy equation,
\begin{eqnarray}
0 &=&-k_n S_{n+1} - b k_nS_n + (1+b) k_{n-2}S_{n-1} \nonumber\\
&-&  k_{n} \langle u_n^*B^*_{n+1} B_{n+2}\rangle \ .\label{slope1}
\end{eqnarray}
When the amplitude of the polymer goes to zero ($B_n\to 0~\forall
n$) the only solution is the well known scaling law $S_n\propto
k_n^{-1}$. However the last term in (\ref{slope1}) forces now a
tilt in the spectrum. Its sign is exactly such that $S_{n-1}$ has
to increase compared to $S_n$ and respectively $S_{n+1}$. Of
course, for $k_n\ll k_c$ the effect of the $B_n$ field on the
$u$-spectrum is again negligible, and therefore the spectral
slope will settle back to the Sabra value. However if the tilt in
the vicinity of $k_c$ results in crossing the Sabra spectrum we
would have a whole spectral range where the energy is higher.

\begin{figure}
\centering
\includegraphics[width=.5\textwidth]{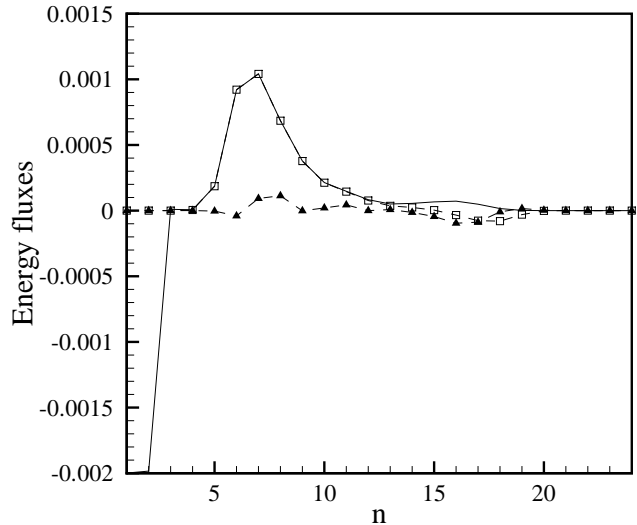}
\caption{Energy fluxes for the SabraP model. The continuous line
corresponds to the fluxes of the kinetic energy $\Psi^{(n)}_1$.
The squares correspond to $-\Psi^{(n)}_2$. The dashed line with
black triangles is $\Psi^{(n)}_3$} \label{signterm}
\end{figure}

We therefore conclude that the existence of drag reduction depends
rather heavily on the sign of the energy transfer at scales close
to $k_c$.  To check the sign directly in the numerics and thus to
substantiate the existence of the tilt we return to the equations
of motion and write
 \begin{eqnarray}
\!\!\! \frac{d}{dt} |u_n|^2 \!\!&=&\!\!
\Psi^{(n)}_1(u,u,u) - \Psi^{(n)}_2(u,B,B) -\nu k_n^2|u_n|^2 \ , \nonumber\\
\label{Bflux} \! \! \!\frac{d}{dt} |B_n|^2 \!\!&=&
\!\!\Psi^{(n)}_3(u,B,B) + \Psi^{(n)}_2(u,B,B)
\!-\!\frac{1}{\tau}|B_n|^2 \ ,
\end{eqnarray}
where the term $\Psi_1^{(n)}(u,u,u)$ represents the kinetic energy
flux of the field, while $\Psi^{(n)}_2(u,B,B)$ is the energy flux
going from the velocity field to the polymer field. Finally,
 term $\Psi^{(n)}_3(u,B,B)$ is the flux of energy of the polymer field
due to the transport of the velocity field. Figure
(\ref{signterm}) shows $\Psi^{(n)}_1$, $-\Psi^{(n)}_2$ and
$\Psi^{(n)}_3$ for $\tau=0.4$ and $\nu = 10^6$, the same
parameters of figure (\ref{logspectrum}). It is important to
observe that $\Psi^{(n)}_1$ becomes positive for $n>n_c$. For a
given $n$ the term $\Psi^{(n)}_1$ can be written as $\Psi^{(n)}_1
= L_n - S_n$ where $L_n$ is the amount of energy flux given from
the large scale to scale $k_n$ and $S_n$ is the amount of energy
flux given from scale $k_n$ to smaller scales. It follows that
when the energy flux is constant $L_n = S_n$ and therefore
$\Psi^{(n)}_1(n) = 0$. On the other hand, a positive value of
$\Psi^{(n)}_1(n)$ implies that $L_n
>S_n$. This is exactly what is shown in figure (\ref{signterm}).
The imbalance of the energy flux $\Psi^{(n)}_1$ is compensated by
the flux of energy from $u_n$ to $B_n$, given by the term
$-\Psi^{(n)}_2$. It is interesting to observe that the last term
in the balance equation, namely $\Psi^{(n)}_3$, is rather small,
i.e. the effect of an energy cascade of the polymer is rather
weak.

\subsection{Discussion}
\label{drag} 
\begin{figure}
\centering
\includegraphics[width=.5\textwidth]{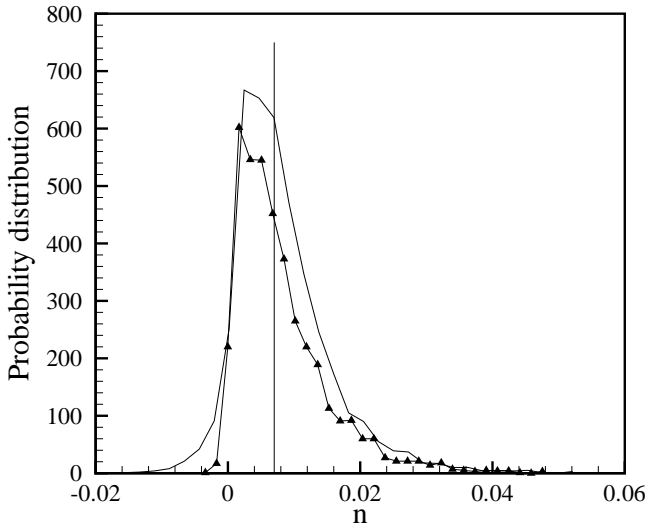}
\caption{Probability distribution functions of $W_L$ for Sabra
model (line with black triangles) and SabraP model (continuous
line).} \label{figprob}
\end{figure}

While we have been able so far to describe a
convincing scenario for drag reduction, we still should explain
the mechanism for the increase of the large scale energy. Since
the field $B_n$ is negligible for small $n$, the average energy
flux per unit time at small $n$ must equal the input work per
unit time at the largest scales. However, {\em the energy flux
does show time and scale fluctuations which could behave
differently for the Sabra and SabraP models}. More specifically
let us consider the quantity $G_n$ defined in Subsect.
\ref{typical}. As already discussed, $G_n$ represents the energy
flux at scale $k_n$ due to both the non linear terms in the
velocity field {\em and} the non linear term in the $B_n$ field.
In terms of $G_n$, we can build a large scale energy flux $W_L =
G_2+G_3+G_4$ which represents the full amount of energy flux
across the largest scales, namely across $k$-vectors $k_n<k_c$,
for which the average energy flux is invariant to changing Sabra
to SabraP. The definition of $W_L$ is such that $W_L
> 0$ means an energy flux from large scales to small scales. In figure
(\ref{figprob}), we show the probability distribution of $W_L$ for
both models, with numerical parameters $\nu = 10^{-5}$ and $\tau=
0.4$. The vertical line in the figure indicates the average value,
which, as expected is invariant. As one can observe, the two
probability distributions show a substantial difference for {\em
negative} values of $W_L$: in the SabraP model one can have larger
and more frequent negative values of the energy flux. This can
happen even if the instantaneous value of the total energy per
unit time obtained by the polymer field from the velocity field
is always positive! Qualitatively this means that while the
velocity field is always forcing the polymer field, at very large
scales the flux can, from time to time, be reversed, and the
polymer field forces there the velocity. This is how the
amplitude of the energy spectrum is being increased on the
average.

\begin{figure}
\centering
\includegraphics[width=.5\textwidth]{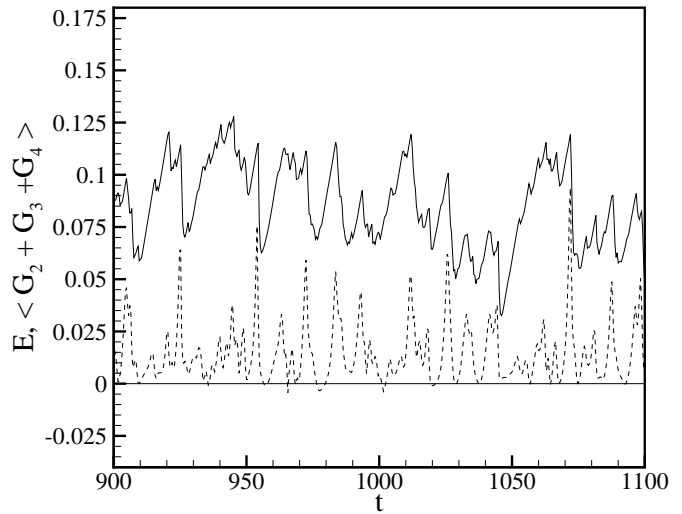}
\caption{Time series of the kinetic energy (continuous line) and
$2\times W_L$ (dotted line) for the Sabra model.} \label{timens}
\end{figure}

\begin{figure}
\centering
\includegraphics[width=.5\textwidth]{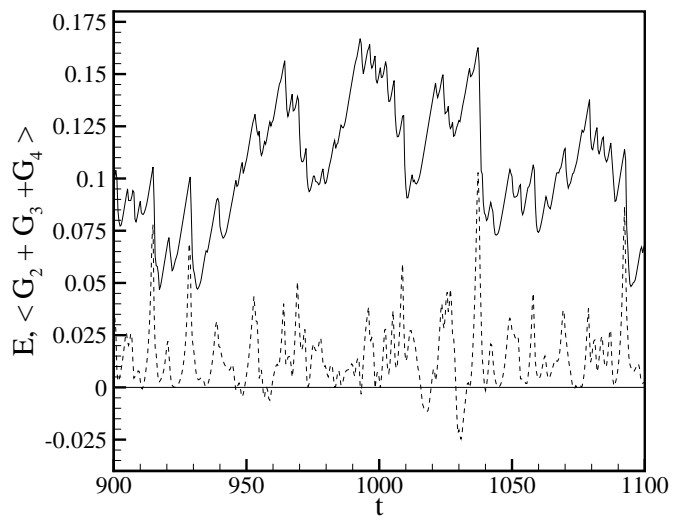}
\caption{Time series of the kinetic energy (continuous line) and
$2\times W_L$ (dotted line) for the SabraP model.}
\label{timepoli}
\end{figure}

Clearly, this mechanism could not work unless the ``forcing" by
the polymer field acted in phase with the growing  kinetic
energy. In order to clarify this further we present two
time-series of the kinetic energy and $W_L$,  in Fig.
(\ref{timens}) for the Sabra model, and in Fig. (\ref{timepoli})
for the SabraP model.
A close inspection of the figures shows that the reverse of the
energy flux $W_L$ occurs exactly during the growing phase of the
kinetic energy, leading therefore to a larger value of the
instantaneous kinetic energy. This {\em in phase} mechanism is
responsible for drag reduction. Note that this mechanism strongly
depends on the large scale dynamics and the value of $k_c$. For
$k_n$ larger than $k_c$ no significant difference in the
statistical behaviour of the energy flux is observed. However,
the amount of energy forcing, due to the polymer at large scale,
can depend on the Reynolds number, at least in the SabraP model.
If this is the case, then drag reduction should depend on the
Reynolds number only through the two relevant scales appearing in
the systems, namely $k_c$ and $\lambda_T$, the latter being the
Taylor microscale
\begin{equation}
\lambda_T \equiv \sqrt{\frac{E_u}{\sum k_n^2 |u_n|^2}} \ ,
\end{equation}
which also depends on $\tau$ in the SabraP model. Because the
drag is a dimensionless quantity, we argue that the only way in
which the Reynolds number may appear in the drag reduction is by
means of the dimensionless quantity $\mu = k_c \lambda_T(\tau)$.

There is a simple argument, proposed in the next section, which
explains why at large Reynolds numbers one may observe the same
qualitative mechanism, i.e. drag reduction, with smaller effects
on the kinetic energy. As a matter of fact, the numerical results
show that drag reduction reaches its maximum for $\mu \sim 1$

\section{Predictions of the theoretical mechanism}

\begin{figure}
\centering
\includegraphics[width=.5\textwidth]{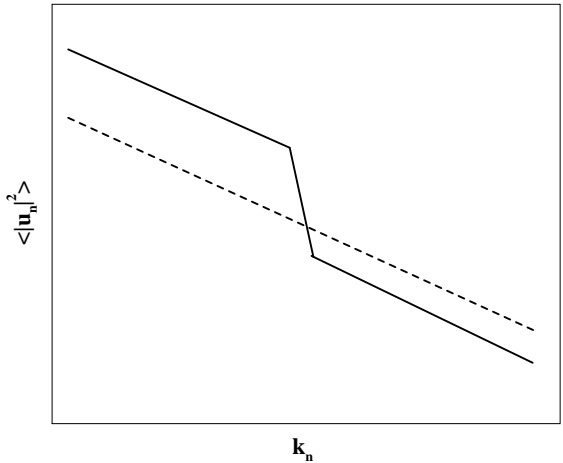}
\caption{Schematic of the effect of polymers in drag reduction on
the turbulence energy spectrum. Dotted line: neat fluid. Solid
line: polymeric solution. The spectral slope is unchanged for
large and small scales, while at scale $k_c$ there is a
significant upward tilt. } \label{cartoongood}
\end{figure}

We summarize the mechanism of drag reduction using the cartoon shown
in Fig. \ref{cartoongood}.
The tilt in the spectrum occurs in the vicinity of $k_c$, with
the asymptotic slope for $k_n\ll k_c$ and $k_n\gg k_c$ remaining
essentially unchanged. With such a spectrum the two inequalities
(\ref{diss-ineq}) and (\ref{drag-red}) are obviously obeyed.

\begin{figure}
\centering
\includegraphics[width=.5\textwidth]{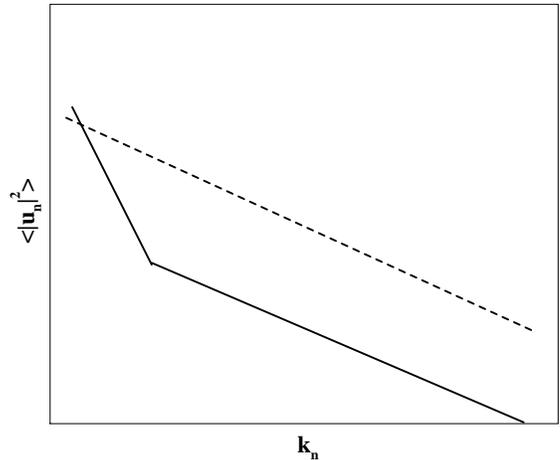}
\caption{Schematic of the turbulence energy spectrum when the
polymer relaxation time scale is too slow, and the Reynolds
number is too large.} \label{cartoonfar}
\end{figure}

\begin{figure}
\centering
\includegraphics[width=.5\textwidth]{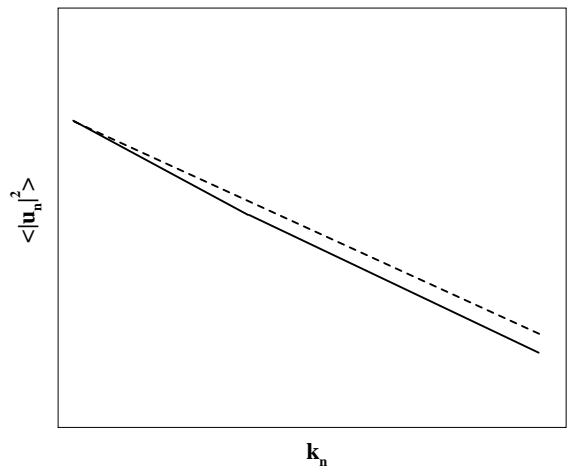}
\caption{The relaxation time $\tau$ is to low.}
\label{cartoonnear}
\end{figure}

The difference in the spectra for $k_n \gg k_c$ is determined
predominantly by Eq. (\ref{disp-diff}). This equation predicts
that this difference will be greatly increased when the Reynolds
number is increased (i.e. when $\nu\to 0$), see Fig.
\ref{cartoonfar}. Of course, if this happens we can lose the whole
effect of drag reduction, since the amount of tilt at $k_c$ is
basically independent of $\nu$. We need to maintain the spectral
difference small enough for the tilt to effect a crossing of the
spectrum of SabraP and Sabra. Also the position of $k_c$ is
important. If we reduce $k_c$ (i.e. increase $\tau$) the tilt is
too far to the left and therefore it will fail to increase the
energy. In fact it can be drag enhancing. The combined effect of
decreasing $\nu$ and increasing $\tau$ is shown in Fig.
\ref{cartoonfar}.
Needless to say, also if we decrease $\tau$ too much we may lose
drag reduction since the tilt will be pushed to the irrelevant
dissipative range where no energy containing modes exist. Also, if
$\tau$ becomes too low, $B_n$ becomes smaller, and the amount of
tilt is decreased, as can be seen directly from Eq.
(\ref{slope1}). Although decreasing the field $B_n$ brings the
spectra closer together in the large $k_n$ regime, the tilt may
not suffice to reduce the drag. Such a situation is shown
schematically in Fig. \ref{cartoonnear}.
Actually, using the language introduced in the previous section
and the above discussions, we are able to give another argument
to understand how drag reduction could depend on the Reynolds
number. As previously said, the relevant dimensionless number in
the system is $\mu = k_c \lambda_T$. If $\tau \rightarrow \infty$
then $k_c \rightarrow 0$ and we know that drag reduction must be
inhibited. It follows that for $\mu \rightarrow 0$ we cannot
observe drag reduction. For fixed $k_c$ and increasing Reynolds,
$\lambda_T(\tau)$ decreases as well, although not necessarily as
$Re^{-1/2}$, where $Re$ is the Reynolds number. Then, for fixed
$\tau$ and increasing Reynolds number we should observe a
decreasing effect of drag reduction, as observed in our numerical
simulation. The same reasoning can be applied to get information
for small Reynolds numbers, as the following argument shows. For
$\tau \rightarrow 0$ we have already shown that no drag reduction
is possible simply because $k_c \rightarrow \infty$. This is
equivalent to say that when $\mu$ becomes too large there cannot
be any drag reduction. It follows that for small $Re$, i.e. for
large $\lambda_T$, drag reduction disappears.

\section{Conclusions and Discussions}
\label{conclusions}

In this paper we discussed several points concerning
the possible formulation of a theory for drag reduction in turbulent
flow with dilute polymer. It is worthwhile,
therefore, to review the main points.

A) We have introduced a shell model resembling the dynamical
properties of the FENE-P equations. Beside any theoretical
considerations, the model shows drag reduction in a way close to
what already observed in the numerical simulations of the FENE-P
model. The implications of this result is that one need not focus
on boundary effects or dynamical properties of coherent structure
in order to capture the basic physics of drag reduction.

B) There exist a relevant scale in the system, $k_c$ defined by
the so called 'time criterion', i.e. $u(k_c)k_c \sim \tau^{-1}$.
In the vicinity of this scale there is a tilt in the spectrum
which causes a crossing of the SabraP velocity spectrum above the
Sabra spectrum for $k_n<k_c$. This in its turn means an increase
in the kinetic energy at large scales.  Drag reduction can be
physically understood in terms of the energy exchanges between
the velocity field and the polymer field for $k_n \sim k_c$. We
have succeeded in proposing a coherent picture, based on the
equation of motions, for the dynamics which is in close agreement
with the numerical results.

C) Drag reduction is a property of large scale flow and its
dynamics. This implies that a quantitative description of drag
reduction must depend on the details of the flow, the forcing
mechanism as well as the Reynolds numbers. Although a general
qualitative mechanism should occur in all drag reduction flow,
the amount of the drag reduction itself depends on how much energy
is {\em intermittently} given to large scale velocity. Thus large
scale fluctuations are important for a quantitative theory.

D) Drag reduction by no means could be reduced to the dynamics at
the dissipation scale. Although drag reduction could be Reynolds
dependent, drag reduction cannot be reduced to a simple increase
of the dissipation length. Actually, the dissipation scale does
not seem to be affected by drag reduction.

\begin{acknowledgments}
We thank Carlo Casciola, Victor L'vov and Renzo Piva for many
useful discussions. This work had been supported in part by
European Commission under a TMR grant ``Non-ideal Turbulence" and
The Minerva Foundation, Munich, Germany.
\end{acknowledgments}

\end{document}